\documentclass{article}

\usepackage{arxiv}

\usepackage[utf8]{inputenc} % allow utf-8 input
\usepackage[T1]{fontenc}    % use 8-bit T1 fonts
\usepackage{hyperref}       % hyperlinks
\usepackage{url}            % simple URL typesetting
\usepackage{booktabs}       % professional-quality tables
\usepackage{amsfonts}       % blackboard math symbols
\usepackage{nicefrac}       % compact symbols for 1/2, etc.
\usepackage{microtype}      % microtypography
\usepackage{lipsum}		% Can be removed after putting your text content
\usepackage{graphicx}
\usepackage{natbib}
\usepackage{doi}

\usepackage{amsmath}
\usepackage{bm}
\usepackage{color}

\title{Full Information Linked ICA: addressing missing data problem in multimodal fusion \\ \large{For the Alzheimer’s Disease Neuroimaging Initiative} \thanks{Data used in the preparation of this article were obtained from the Alzheimer's Disease Neuroimaging Initiative (ADNI) database (adni.loni.usc.edu). As such, the investigators within the ADNI contributed to the design and implementation of ADNI and/or provided data but did not participate in the analysis or writing of this report. A complete listing of ADNI investigators can be found at: \url{http://adni.loni.usc.edu/wp-content/uploads/how_to_apply/ADNI_Acknowledgement_List.pdf}}
}

\date{} 					% Or removing it

\author{ \href{https://orcid.org/0000-0002-8635-2669}{\includegraphics[scale=0.06]{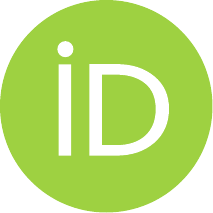}\hspace{1mm}Ruiyang Li} \\
	Department of Biostatistics\\
	Columbia University\\
	New York, NY \\
	\And
	F. DuBois Bowman \\
        School of Public Health \\
	University of Michigan \\
	Ann Arbor, MI \\
	\And
	\href{https://orcid.org/0000-0003-3177-6357}{\includegraphics[scale=0.06]{orcid.pdf}\hspace{1mm}Seonjoo Lee} \thanks{To whom correspondence should be addressed} \\
	Department of Biostatistics and Psychiatry\\
	Columbia University, New York State Psychiatric Institute\\
	New York, NY \\
	\texttt{seonjoo.lee@nyspi.columbia.edu} \\
}

\renewcommand{\shorttitle}{Full Information Linked ICA: addressing missing data problem in multimodal fusion}

\hypersetup{
pdftitle={Full Information Linked ICA: addressing missing data problem in multimodal fusion},
pdfsubject={stat.ME, stat.CO, stat.AP},
pdfauthor={Ruiyang Li, F. DuBois Bowman, Seonjoo Lee},
pdfkeywords={full information maximum likelihood, linked independent component analysis, missing data, multimodal fusion},
}

\begin{document}
\maketitle

\begin{abstract}
Recent advances in multimodal imaging acquisition techniques have allowed us to measure different aspects of brain structure and function. Multimodal fusion, such as linked independent component analysis (LICA), is popularly used to integrate complementary information. However, it has suffered from missing data, commonly occurring in neuroimaging data. Therefore, in this paper, we propose a Full Information LICA algorithm (FI-LICA) to handle the missing data problem during multimodal fusion under the LICA framework. Built upon complete cases, our method employs the principle of full information and utilizes all available information to recover the missing latent information. Our simulation experiments showed the ideal performance of FI-LICA compared to current practices. Further, we applied FI-LICA to multimodal data from the Alzheimer's Disease Neuroimaging Initiative (ADNI) study, showcasing better performance in classifying current diagnosis and in predicting the AD transition of participants with mild cognitive impairment (MCI), thereby highlighting the practical utility of our proposed method.
\end{abstract}

% keywords can be removed

\section{Introduction}
\label{sec:intro}

\begin{figure}
\begin{center}
\includegraphics[scale=0.7]{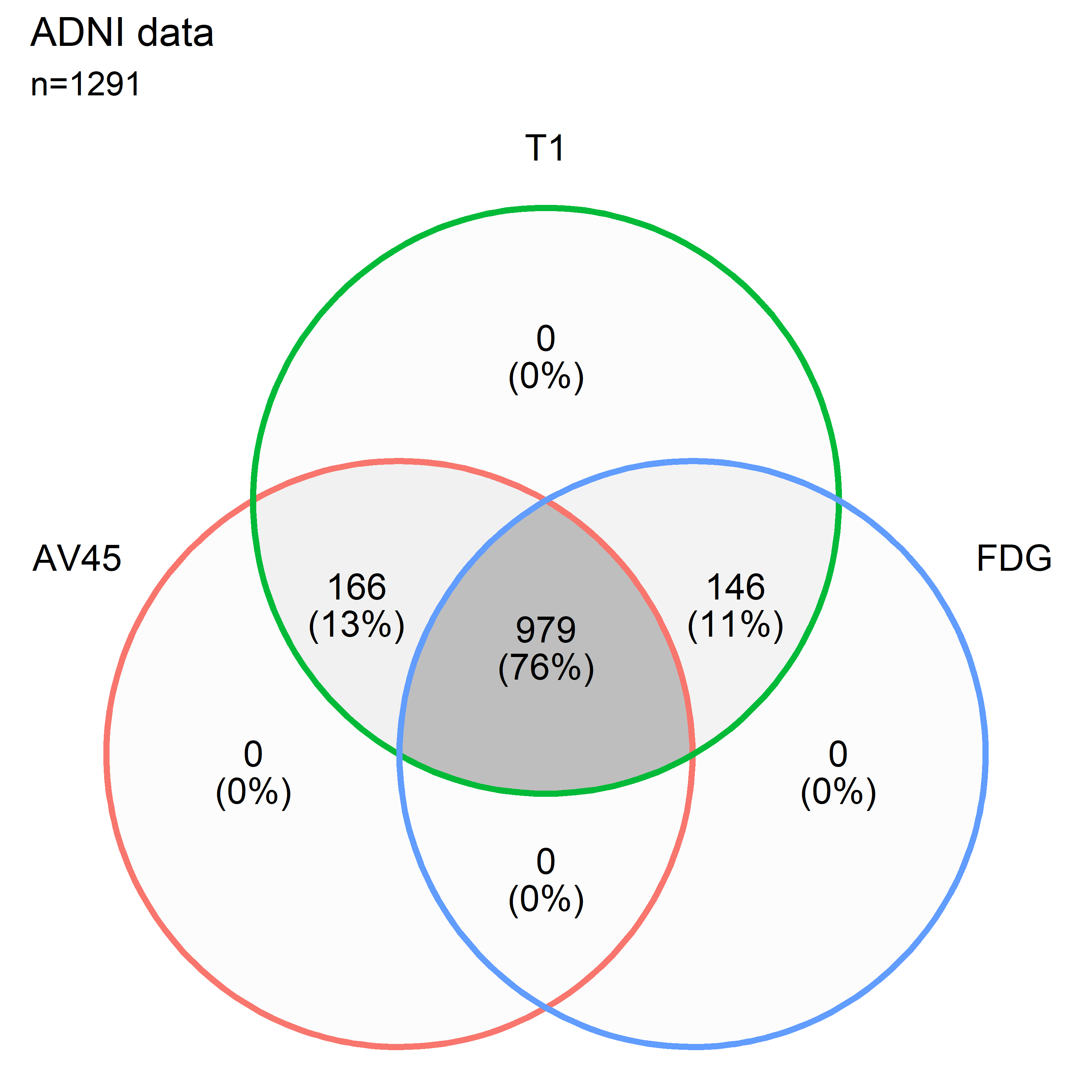}
\end{center}
\caption{The Venn diagram of modality availability (T1 MRI, AV45 PET, and FDG PET) in ADNI dataset at baseline. 
\label{fig:adni_mod_venn_diagram}}
\end{figure}

Alzheimer’s disease (AD) is an irreversible neurodegenerative brain disorder that progressively compromises cognitive functioning and behavioral abilities \citep{yu_mapping_2022}. In AD studies, data from different imaging modalities, such as Magnetic Resonance Imaging (MRI) and Positron Emission Tomography (PET) imaging, is typically collected to derive important biomarkers for disease diagnosis and progression. 
To understand the cross-information from different modalities in a more unified way, it becomes increasingly necessary to analyze data from different modalities jointly \citep{lahat_multimodal_2015, calhoun_multimodal_2016}, which introduces the concept of multimodal fusion. 
However, such information integration is impeded by missing data, a common occurrence in neuroimaging data for various reasons including quality check failure, study design, or the absence of specific imaging sessions. Thus, handling missing data during the fusion of multiple modalities is not trivial in AD studies because of distinct causes and patterns of missing data. 
Take as an illustrative example the data from the Alzheimer’s Disease Neuroimaging Initiative (ADNI), a longitudinal multicenter study designed for the early detection and tracking of Alzheimer’s disease (Section \ref{sec:analysis}). Different imaging data was collected in the ADNI study, such as T1-weighted MRI, Florbetapir (AV45) PET imaging, and fluorodeoxyglucose (FDG) PET imaging. There were 1291 subjects who had no missing T1 data but missing AV45 or FDG data at baseline. Figure \ref{fig:adni_mod_venn_diagram} displays the Venn diagram of the sample size within the three modalities. It can be observed that a considerable portion of data (24\%) would be lost if only complete data across all three modalities were considered. The sample size would decrease further if more modalities were taken into consideration, given the complex data structure among modalities. 
Therefore, it is important to properly address the missing data problem in multimodal fusion to ensure accurate and reliable results. 

Linked Independent Component Analysis (LICA; \citealp{groves_linked_2011}) has been widely used for multimodal fusion (e.g., in \cite{liu_multimodal_2021, maglanoc_multimodal_2020}) because of the ease of interpretation. It decomposes data from multiple modalities into a pre-specified number of independent components and obtains the shared component-wise subject loadings across modalities. Specifically, under the LICA framework, data from each modality $k=1, \cdots, K$ can be written as 
\begin{equation} \label{eqn:lica}
\mathbf{Y}^{(k)} = {\mathbf{X}^{(k)}_\mathbf{W}} \mathbf{H} + \mathbf{E}^{(k)},
\end{equation}
where $\mathbf{Y}^{(k)}$ is the data matrix in modality $k$ (with each row representing a voxel in modality $k$ and each column representing a subject), $\mathbf{X}^{(k)}_\mathbf{W} = \mathbf{X}^{(k)} \mathbf{W}^{(k)}$ is the weighted spatial map in modality $k$ (structured as a matrix where each row represents a voxel in modality $k$ and each column represents an independent component), 
$\mathbf{X}^{(k)}$ is the spatial map in $k$, $\mathbf{W}^{(k)}_{L \times L} = diag(W_1, \cdots, W_L)$ is the weight matrix in $k$, $L$ is the number of components, $\mathbf{H}$ is the latent space (subject-course) matrix shared across all modalities (with each row representing an independent component and each column representing a subject), and $\mathbf{E}^{(k)}$ is the noise matrix in modality $k$. Note that the shared latent component-wise subject loading matrix $\mathbf{H}$ is often of interest for future analysis. 
However, LICA cannot handle missing data very well. 
Currently, there are two ways to handle missing data in LICA. One is the complete case analysis (i.e., only analyzing data from subjects present in all modalities), and the other is the zero-filling method (i.e., replacing missing values with zeros before running LICA). Both of them have their own limitations. The effective sample size in the complete case analysis can drop dramatically when there is a large proportion of missing data, especially when the missing structure among modalities is complicated. As a result, the complete case analysis may discard important underlying information and lead to biased results. In addition, as the shared subject loadings are often of interest in LICA, one notable limitation of the complete case analysis is that the latent information is not available for the excluded subjects and cannot be recovered. Furthermore, the zero-filling method will make the LICA algorithm unstable and thus yield inaccurate estimations.  

Besides LICA, there are also a number of multimodal fusion methods available, including methods based on ICA (e.g., joint ICA \citep{calhoun_method_2006}, parallel ICA \citep{liu_parallel_2007}, coefficient-constrained ICA \citep{sui_method_2009}), methods based on canonical correlation analysis (CCA; e.g., multimodal CCA \citep{correa_canonical_2008}, multiset CCA \citep{correa_multi-set_2010}), methods based on partial least squares (PLS; e.g., multi-way PLS \citep{martinez-montes_concurrent_2004}, multimodal PLS \citep{chen_linking_2009}), and methods based on machine learning and deep learning (a detailed review has been summarized in \citet{stahlschmidt_multimodal_2022}). Each of them has its own benefits and purposes, but the missing data problem still has not been well addressed. Commonly used methods to handle missingness in multimodal fusion include complete data analysis (e.g., in \citealp{simidjievski_variational_2019, tong_deep_2020, lu_ai-based_2021, hira_integrated_2021}), one-time filling for the missing data (e.g., mean-filling (in \citealp{xie_group_2019, zhang_integrated_2019}), median-filling (in \citealp{guo_deep_2020,chai_integrating_2021}), zero-filling (in \citealp{zhang_combining_2020,venugopalan_multimodal_2021,vale-silva_long-term_2021}), filling based on nearest neighbors (in \citealp{chaudhary_deep_2018,sun_multimodal_2019,xu_hierarchical_2019,lv_survival_2020,yu_model_2020,albaradei_metacancer_2021})) with or without a pre-filtering step, and deep learning (e.g., in \citealp{cheerla_deep_2019,shen_brain_2019,sun_semi-supervised_2021,tran_missing_2017,zhou_missing_2022}). 
The same limitation persists with complete data analysis as aforementioned in LICA that important information may be discarded and biased results may be obtained. One-time filling/imputation is prone to yield biased and inaccurate results. Although methods based on deep learning can computationally handle missingness, it is difficult to make inferences from these methods. Therefore, we would like to address the missing data problem in multimodal fusion and we will formulate our method in particular under the framework of LICA.

In classical statistical analysis literature, multiple imputation (MI; \citet{rubin_multiple_1987}) or full information maximum likelihood (FIML; \citet{anderson_maximum_1957}) is commonly used to address the issue of missingness. However, under the context of multimodal fusion, it is not straightforward how MI can be implemented. MI can be viewed as a three-step process -- i) generating several imputed datasets (by imputing missing values in one data using the other available data), ii) performing the target analysis on each of the imputed datasets, and iii) pooling results together to obtain the final results. Under the LICA framework, when the missing data in one modality is imputed using the other modalities in step i), unwanted correlation may be imposed between modalities. This in turn may affect the fusing results in step ii). Further, since the order of the estimated independent components from LICA is random, it is not straightforward how the results can be aggregated in step iii). As such, to address the missing data problem in LICA, one of the plausible ways currently is to use the FIML.
Therefore, in this paper, we utilized the principle of full information and proposed the Full Information Linked ICA algorithm (FI-LICA) to handle the missing data problem in multimodal fusion under the LICA framework. 

In this paper, we first introduced our proposed algorithm in Section \ref{sec:meth}. Then, we presented the simulation results in Section \ref{sec:sim}. Finally, we applied our method to the real-world human brain multimodal imaging data and illustrated its utility in Section \ref{sec:analysis}.

\section{Methods}
\label{sec:meth}

\begin{figure}
\begin{center}
\includegraphics[scale=0.5]{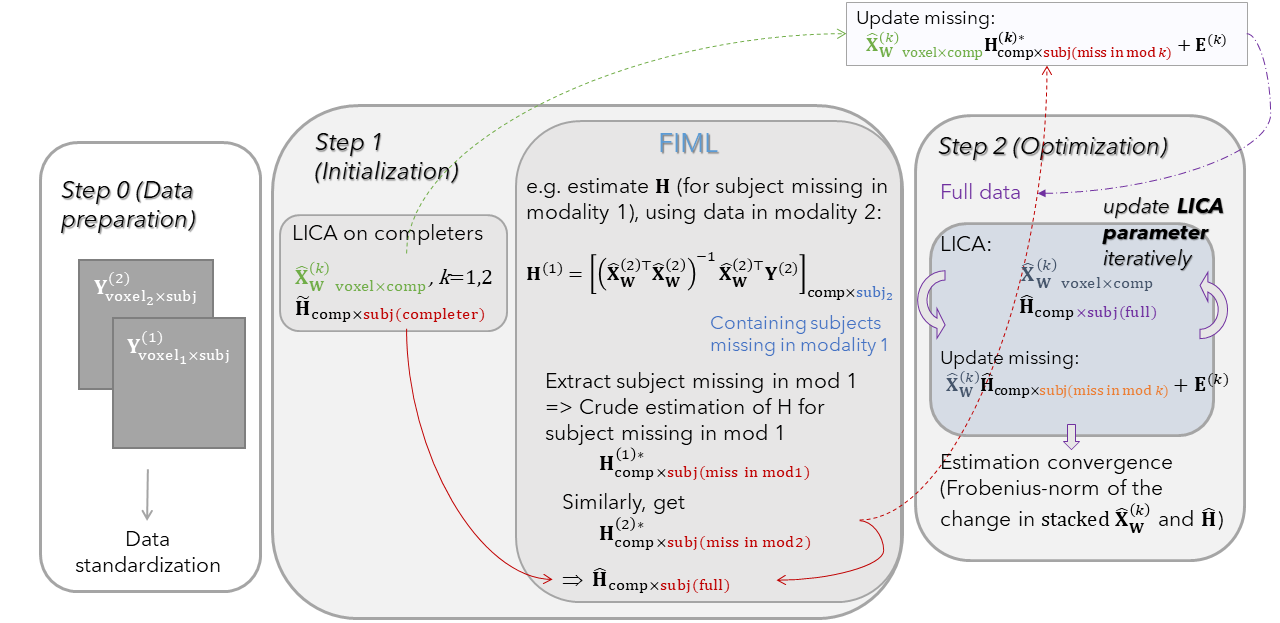}
\end{center}
\caption{The illustrative diagram of FI-LICA with two modalities
\label{fig:diagram}}
\end{figure}

We proposed a Full Information Linked ICA algorithm (FI-LICA) to deal with the missing data problem in multimodal fusion under the LICA framework. Our method is built upon the complete cases and incorporates all the available information to recover the missing latent information. Our algorithm is summarized in Table \ref{tab:tabone}, with an illustrative diagram of two modalities in Figure \ref{fig:diagram}. Details are described below.

\textit{Step 0: Data preparation}

First, data from each modality is standardized based on the availability of subjects. Specifically, for each voxel, the original values are standardized by subtracting their mean and dividing by their root mean square.

\textit{Step 1: Initialization}

In the initialization step, we first focus on the complete cases present in all modalities and estimate the weighted $\mathbf{X}$ for each modality $k$ (i.e., $\hat{\mathbf{X}}^{(k)}_\mathbf{W}$) and the shared subject loading matrix (i.e., $\Tilde{\mathbf{H}}$) using LICA. Then, given the estimation, we recover the missing information of $\mathbf{H}$ for subjects with missing data by leveraging all the available data. 
Specifically, we obtain the crude estimation of $\mathbf{H}$ for subjects with missing data in each modality using data from the rest of the modalities, since $\mathbf{H}$ is shared across modalities and since subjects missing in one modality should have data in the other modalities (otherwise they should not have been considered as they have no available data). Data in modalities except for $k$ can be represented as $\mathbf{Y}^{(-k)} = \mathbf{X}^{(-k)}_\mathbf{W} \mathbf{H} + \mathbf{E}^{(-k)}$, where $\mathbf{Y}^{(-k)}$ is the stacked data from modalities except for $k$, $\mathbf{X}^{(-k)}_\mathbf{W}$ is the stacked weighted spatial maps from modalities except for $k$, and $\mathbf{E}^{(-k)}$ is the noise matrix. The crude estimation of $\mathbf{H}$ for the missing subjects in modality $k$, denoted as $\mathbf{H}^{(k)\ast}$, can be computed from $\mathbf{H}^{(k)} = (\hat{\mathbf{X}}^{(-k)\top}_{\mathbf{W}} \hat{\mathbf{X}}^{(-k)}_{\mathbf{W}})^{-1} \hat{\mathbf{X}}^{(-k)\top}_{\mathbf{W}} \mathbf{Y}^{(-k)}$, where pseudoinverse is used, by focusing on the columns that correspond to the missing subjects in modality $k$. Then, all the crude estimation of $\mathbf{H}$ for the missing subjects (i.e., $\mathbf{H}^{(k)\ast}$s) can be column-wisely combined to the initial estimation of $\mathbf{H}$ from the complete cases (i.e., $\Tilde{\mathbf{H}}$) based on the order of subjects to obtain the crude estimation of $\mathbf{H}$ for all subjects, denoted as $\hat{\mathbf{H}}$. 

\textit{Step 2: Optimization}

To make the estimations more accurate and stable, we employ an iterative process to update the estimations until convergence. 

Now that from Step 1, both the estimated weighted $\mathbf{X}$ (i.e., $\hat{\mathbf{X}}_\mathbf{W}^{(k)}$) and the crude estimation of $\mathbf{H}$ (i.e., $\mathbf{H}^{(k)\ast}$) for subjects missing in modality $k$ are available, the missing data in each modality $k$ can be updated using $\hat{\mathbf{Y}}^{(k)}_{miss} = \hat{\mathbf{X}}_{\mathbf{W}}^{(k)} \mathbf{H}^{(k)\ast}+\mathbf{E}^{(k)}$. 
This gives the full data for all subjects. 

Then, in this iteration step, we keep refitting LICA and updating the missing values given the estimations from LICA until the estimations converge. 
Missing values in each modality $k$ are updated based on $\hat{\mathbf{Y}}^{(k)}_{miss} = \hat{\mathbf{X}}^{(k)}_\mathbf{W} \hat{\mathbf{H}} + \mathbf{E}^{(k)}$, focusing on subjects missing in modality $k$. The Frobenius-norm of the change in stacked weighted $\mathbf{X}$ and in $\mathbf{H}$ are used to check if the estimations converge. 

In summary, Step 1 is the process in which we get the crude estimation of $\mathbf{H}$ for all subjects, and Step 2 is the LICA parameter optimization process in which we refine the crude estimations.

\begin{table}
\footnotesize
\caption{FI-LICA Algorithm Summary \label{tab:tabone}}
\begin{center}
\begin{tabular}{p{15.5cm}}
\hline
Input: multimodal data $\mathbf{Y}^{(k)}, k=1, \cdots, K$ \\\hline
\textbf{Step 0: Data preparation} \\
 \ \ \ \ \ \ Standardize data in each modality based on subjects' availability \\
 
 %\textbf{ } \\
 
 \textbf{Step 1: Initialization} \\
 \ \ \ \ \ \ \textbf{1)} Estimate weighted $\mathbf{X}$ for each modality $k$ ($\hat{\mathbf{X}}_\mathbf{W}^{(k)}$) and shared H ($\Tilde{\mathbf{H}}$) using LICA on completers \\
 \ \ \ \ \ \ \textbf{2)} Obtain crude estimation of $\mathbf{H}$ for subjects with missing in $k$\\
 \ \ \ \ \ \ \ \ \ - Compute $\mathbf{H}^{(k)} = (\hat{\mathbf{X}}^{(-k)\top}_\mathbf{W} \hat{\mathbf{X}}^{(-k)}_\mathbf{W})^{-1} \hat{\mathbf{X}}^{(-k)\top}_\mathbf{W} \mathbf{Y}^{(-k)}$\\
 \ \ \ \ \ \ \ \ \ - Extract columns corresponding to subjects missing in modality $k$, denoted as $\mathbf{H}^{(k)\ast}$ \\
 \ \ \ \ \ \ \textbf{3)} Update missing data in modality $k$ using $\hat{\mathbf{Y}}_{miss}^{(k)} = \hat{\mathbf{X}}_\mathbf{W}^{(k)} \mathbf{H}^{(k)\ast} + \mathbf{E}^{(k)}$ \\
 \ \ \ \ \ \ \textbf{4)} Initial estimation\\
 \ \ \ \ \ \ \ \ \ - Combine all $\mathbf{H}^{(k)\ast}$s to $\Tilde{\mathbf{H}}$ column-wisely based on the order of subjects, denoted as $\hat{\mathbf{H}}[0]$\\
 \ \ \ \ \ \ \ \ \ - Denote $\hat{\mathbf{X}}_\mathbf{W}^{(k)}$ as $\hat{\mathbf{X}}_\mathbf{W}^{(k)}[0]$ \\

 %\textbf{ } \\
 
 \textbf{Step 2: Optimization} \\
 \textit{Now, we have data for all subjects after Step 1.} \\
 %\For{$s = 1, 2, \cdots, S$}{
 For $s = 1, 2, \cdots, S$: \\ 
 \textbf{1)} Refit LICA and get estimated weighted $\mathbf{X}$ for each modality $k$ ($\hat{\mathbf{X}}^{(k)}_\mathbf{W} [s]$) and shared $\mathbf{H}$ ($\hat{\mathbf{H}} [s]$) at $s$ \\
 \textbf{2)} Update missing values using $\hat{\mathbf{Y}}^{(k)}_{miss}[s] = \hat{\mathbf{X}}^{(k)}_\mathbf{W} [s] \hat{\mathbf{H}}[s] + \mathbf{E}^{(k)}[s]$, focusing on subjects with missing \\
 \textbf{3)} Compute Frobenius-norm of change in weighted $\mathbf{X}$ ($||\hat{\mathbf{X}}^{(k)}_\mathbf{W} [s] - \hat{\mathbf{X}}^{(k)}_\mathbf{W} [s-1] ||_{F}$) and in $\mathbf{H}$ ($||\hat{\mathbf{H}} [s] - \hat{\mathbf{H}} [s-1 ||_{F}$) \\
 \textbf{4)} Stop if the estimation converges \\
 %}
\hline
\end{tabular}
\end{center}
\end{table}

\ \ \ \ \ \ \

Our proposed method FI-LICA can be implemented using our FILICA R package, where all the LICA-involved process is implemented using FLICA, a LICA implemented in FSL in MATLAB \citep{groves_linked_2011,smith_advances_2004,groves_benefits_2012,douaud_common_2014}. Detailed information about FLICA can be found at \url{http://fsl.fmrib.ox.ac.uk/fsl/fslwiki/FLICA}. There are two parameters that need to be pre-specified for the LICA estimation -- the number of independent components and the number of iterations. The number of independent components is typically determined based on the study of interest, and a large iteration number is usually expected so that the estimation converges (dF $<$ 0.1). For FI-LICA, we keep Step 1 the same as the complete case analysis and implement Step 2 given the number of FI-LICA updates and the number of LICA iterations. A combination of a large FI-LICA update number and a large LICA iteration number is expected for the convergence of weighted $\mathbf{X}$ and $\mathbf{H}$. 

We notice that the currently available implementation of LICA with FLICA in MATLAB does not guarantee the estimated $\mathbf{H}$ to have a unit standard deviation as claimed in the original paper. To address this issue, we rescale the estimated $\mathbf{H}$ using its standard deviation based on $$\mathbf{Y}^{(k)} = {\mathbf{X}^{(k)}_\mathbf{W}}{\mathbf{H}} + \mathbf{E}^{(k)} = ({\mathbf{X}^{(k)}_\mathbf{W}} \mathbf{D}^{-1}) \ (\mathbf{D}{\mathbf{H}}) + \mathbf{E}^{(k)} = {\mathbf{X}^{(k)}_\mathbf{W}}_{\text{rescaled}} \mathbf{H}_{\text{rescaled}} + \mathbf{E}^{(k)}$$ during the algorithm process, where $\mathbf{D} = diag(1/sd(\mathbf{H}^\top))$ is the diagonal matrix with diagonal entries being the inverse of the standard deviation of each component of $\mathbf{H}$ and $\mathbf{D}^{-1} = diag(sd(\mathbf{H}^\top))$ is the diagonal matrix with diagonal entries being the standard deviation of each component of $\mathbf{H}$. 
Also, in Step 2, we use the estimated (rescaled) $\mathbf{H}$ as the starting $\mathbf{H}$ in LICA, instead of using PCA to initialize $\mathbf{H}$ by default in FLICA.

\section{Simulation}
\label{sec:sim}

In this section, we investigate the performance of our proposed method FI-LICA under two missing data mechanisms. One is the Missing Completely At Random (MCAR) mechanism, where the missingness is due to chance only, and the other is the Missing At Random (MAR) mechanism, where the missingness is related to the observed data \citep{mack_types_2018,little_statistical_1987,rubin_inference_1976}.

\subsection{Data generation}

\subsubsection{Simulation setting 1: MCAR}

\begin{figure}
\begin{center}
\includegraphics[scale=0.75]{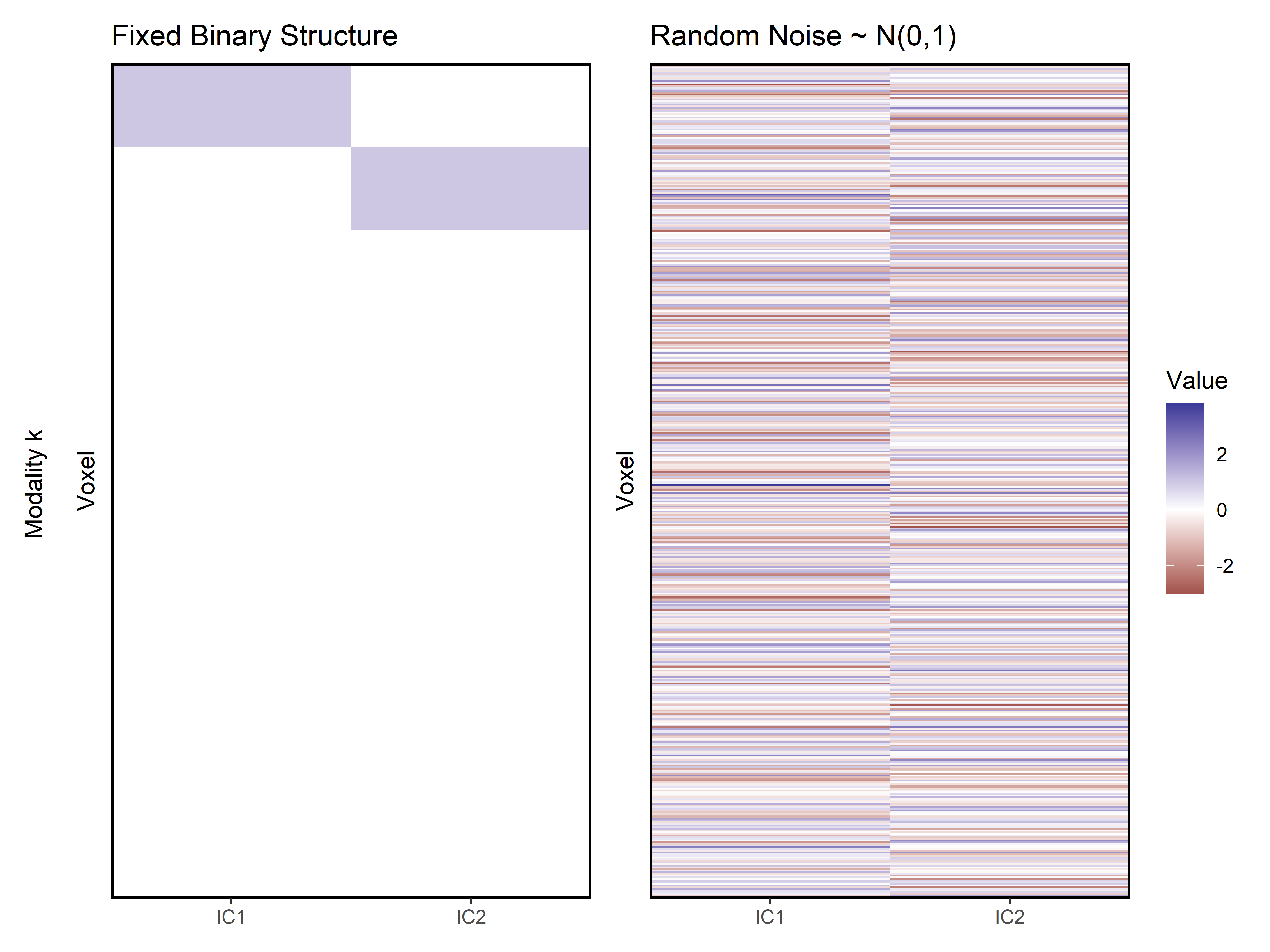}
\end{center}
\caption{The generation of $\mathbf{X}^{(k)}$ with two uncorrelated components, for modality $k = 1, 2$, in all the simulation settings. Each $\mathbf{X}^{(k)}$ was generated as the fixed binary structure plus some random noise from the standard normal distribution. \label{fig:filica_sim_xw}}
\end{figure}

In the simulation setting of MCAR, we started by generating 100 no-missing datasets with 2 modalities and 100 subjects. Modality 1 has 1,000 voxels and modality 2 has 3,000 voxels. The number of true independent components is 2. 
Data in each modality $k = 1, 2$ can be generated using the LICA model in (\ref{eqn:lica}). We let $\mathbf{W}^{(k)} = \mathbf{I}_2$, and $\mathbf{X}^{(k)}_\mathbf{W}$ was reduced to $\mathbf{X}^{(k)}$. To generate $\mathbf{X}^{(k)}_{\text{voxel} \times 2}$ with two uncorrelated components, we imposed a binary structure -- we started by assigning 1 to the first 100 entries of component 1 and the 101 to 200 entries of component 2 and assigning 0 to the rest of entries, and then we added some random noises from $N(0,1)$ to all the entries (this generation process is also shown in Figure \ref{fig:filica_sim_xw}). In addition, we let each element in $\mathbf{H} \sim N(0,1)$ and each element in $\mathbf{E}^{(1)}, \mathbf{E}^{(2)} \sim N(0,1)$. 

To generate the missing datasets, we implemented 5\%, 10\%, and 20\% missing per modality, respectively, on each generated no-missing dataset. The missingness was completely random and thus designed to mimic the mechanism of MCAR. No overlap was made in missing subjects between modalities for each generated missing dataset. 

\subsubsection{Simulation setting 2: MAR (Continuous $\mathbf{H}$)}

Same as in the setting of MCAR, in this MAR setting, we again started with 100 no-missing datasets, each of which has 2 modalities (with 1,000 voxels and 3,000 voxels, respectively), 100 subjects, and 2 true components. 
To mimic the mechanism of MAR, where missingness depends on observed variables, we additionally generated two covariates for each subject. One (i.e., $C_1$) is a continuous variable strongly correlated with the first component of $\mathbf{H}$ (i.e., $H_1$) with a correlation of 0.5. The other (i.e., $C_2$) is a continuous variable mediumly correlated with the second component of $\mathbf{H}$ (i.e., $H_2$) with a correlation of 0.3. That is, we generated $C_1$, $C_2$, $H_1$, and $H_2$ from a multivariate normal distribution, with zero means, unit variances, and a correlation matrix 
$\begin{pmatrix}
  1 & 0 & 0.5 & 0\\
  0 & 1 & 0 & 0.3\\
  0.5 & 0 & 1 & 0\\
  0 & 0.3 & 0 & 1
\end{pmatrix}$. 
Then, $\mathbf{H}$ was generated by combining the standardized $H_1$ and $H_2$ (i.e., ($H_1^\top,H_2^\top)^\top$), to force it to follow $N(0,1)$. 
Again, data in each modality $k$ was generated based on model (\ref{eqn:lica}), with $\mathbf{W}^{(k)} = \mathbf{I}_2$ and uncorrelated $\mathbf{X}^{(k)}_{\text{voxel} \times 2}$ generated as shown in Figure \ref{fig:filica_sim_xw}, and elements in $\mathbf{E}^{(1)}, \mathbf{E}^{(2)} \sim N(0,1)$.

To generate the missing datasets, we again consider implementing 5\%, 10\%, and 20\% missing per modality, respectively, on each generated no-missing dataset. The missingness was generated based on the two covariates and thus designed to mimic the MAR mechanism. The missing probability for each subject was computed based on the logistic model and took the form of $p = 1/(1+e^{-Z})$, where $Z = -0.6 + 0.5 \times C_1 + 1.2 \times C_2$. Subjects with the top largest $p$ were assigned missing in turn for modalities 1 and 2 to achieve the target missing percentage. No overlap was made in missing subjects between modalities for each generated missing dataset.

\subsubsection{Simulation setting 3: MAR (Mixed $\mathbf{H}$)}

In this simulation setting, we aim to investigate the method performance when the distribution assumption of $\mathbf{H}$ is violated. Particularly, we generated $\mathbf{H}$ from a mixture of distributions, rather than a normal distribution assumed under the LICA framework. 

Same as in simulation settings 1 and 2, we started by generating 100 no-missing datasets, each of which has 2 modalities (with 1,000 voxels and 3,000 voxels, respectively), 100 subjects, and 2 true components. 
Similar to the simulation setting 2, we generated two covariates ($C_{1}$ and $C_{2}$) that are related to $\mathbf{H}$. We let $C_1$ be a continuous variable correlated with the first component of $\mathbf{H}$ (i.e., $H_1$). 
However, the difference is that now we let $C_2$ be a binary variable, rather than a continuous variable, related to the second component of $\mathbf{H}$ (i.e., $H_2$). Specifically, $C_1$ and $H_1$ were generated from a bivariate normal distribution, with zero means, unit variance, and a correlation of 0.5. $C_2$ was generated from the Bernoulli distribution with a probability of 0.5, and $H_2$ was generated by multiplying $C_2$ by 0.5 and adding some random noises from the standard normal distribution. Then, $\mathbf{H}$ was generated by combining the standardized $H_1$ and $H_2$ (i.e., ($H_1^\top,H_2^\top)^\top$). 
Again, data in each modality $k$ was generated based on model (\ref{eqn:lica}), with $\mathbf{W}^{(k)} = \mathbf{I}_2$ and uncorrelated $\mathbf{X}^{(k)}_{\text{voxel} \times 2}$ generated as shown in Figure \ref{fig:filica_sim_xw}, and elements in $\mathbf{E}^{(1)}, \mathbf{E}^{(2)} \sim N(0,1)$. 
Missing datasets were generated in the same way as in the simulation setting 2.

\subsection{Method evaluation}

We are interested in evaluating 1) the performance of our proposed method FI-LICA in recovering the spatial maps and the shared subject score matrix, with respect to the truth, and 2) when the missing is MAR, how well the recovered subject score preserves its relationships with the variables that are related to the missing mechanism. 

The best-matching components can be identified based on the stacked weighted spatial maps from all modalities, denoted as $\mathbf{X_W} = \big(\mathbf{X}_\mathbf{W}^{(1)}, \cdots, \mathbf{X}_\mathbf{W}^{(K)} \big)^\top$. Specifically, for each of the two true components, the best-matching component is the estimated $\mathbf{X_W}$ component that is most correlated with the true $\mathbf{X_W}$ component. It follows that the best-matching $\mathbf{X_W}$ and the best-matching $\mathbf{H}$ are the estimated $\mathbf{X_W}$ and the estimated $\mathbf{H}$ focusing on the two best-matching components. For the rest of the sections, we use the best-matching $H_1$ to represent the best-matching $\mathbf{H}$ that corresponds to the first component of true $\mathbf{H}$ and the best-matching $H_2$ to represent the best-matching $\mathbf{H}$ that corresponds to the second component of true $\mathbf{H}$. 

To evaluate the estimation performance, we showed the absolute correlations between the best-matching $\mathbf{X_W}$ and the true $\mathbf{X_W}$, and the absolute correlations between the best-matching $\mathbf{H}$ and the true $\mathbf{H}$, separated by the two true components, and we compared the results of FI-LICA with i) the oracle method (i.e., applying LICA to no-missing data), ii) the complete case analysis (completer), and iii) the replacing-missing-with-0 method (replace0). 

Additionally, for the MAR setting with continuous $\mathbf{H}$, we assessed the bias of the correlation between $C_1$ and the best-matching $H_1$ with the truth (0.5), and the bias of the correlation between $C_2$ and the best-matching $H_2$ with the truth (0.3). 
For the MAR setting with mixed $\mathbf{H}$, we evaluated the bias of the correlation between $C_1$ and the best-matching $H_1$ with the truth (0.5), and the bias of the Cohen's d between $C_2$ and the best-matching $H_2$ with the truth (0.5). 
Note that correlations and Cohen's d measurements were adjusted for the signs of the correlations between the corresponding best-matching $\mathbf{H}$ and true $\mathbf{H}$.

%\subsection{Parameter specification}

In all the simulation settings, five components were used for estimation. 1500 iterations were used when LICA was applied to the no-missing data and in complete case analysis. Estimation converged (dF $<$ 0.1) in both cases. For FI-LICA, a combination of 1000 LICA iterations and 20 FI-LICA updates was used. 
However, there are some practical considerations that we are aware of for the replace0 method with the current LICA implementation. In the replace0 analysis, LICA can give unstable estimation (i.e., unstable and divergent dF), which is not desired. In addition, it can stop early especially when the missing data has a large percentage of missing. In cases where LICA stopped early, we reduced the starting iteration number (e.g., 1500) by 25\% (round up to the nearest integer if the reduction does not give a whole number) and reran LICA until it worked.

\subsection{Results}

\subsubsection{Simulation setting 1: MCAR}

\begin{figure}
\begin{center}
\includegraphics{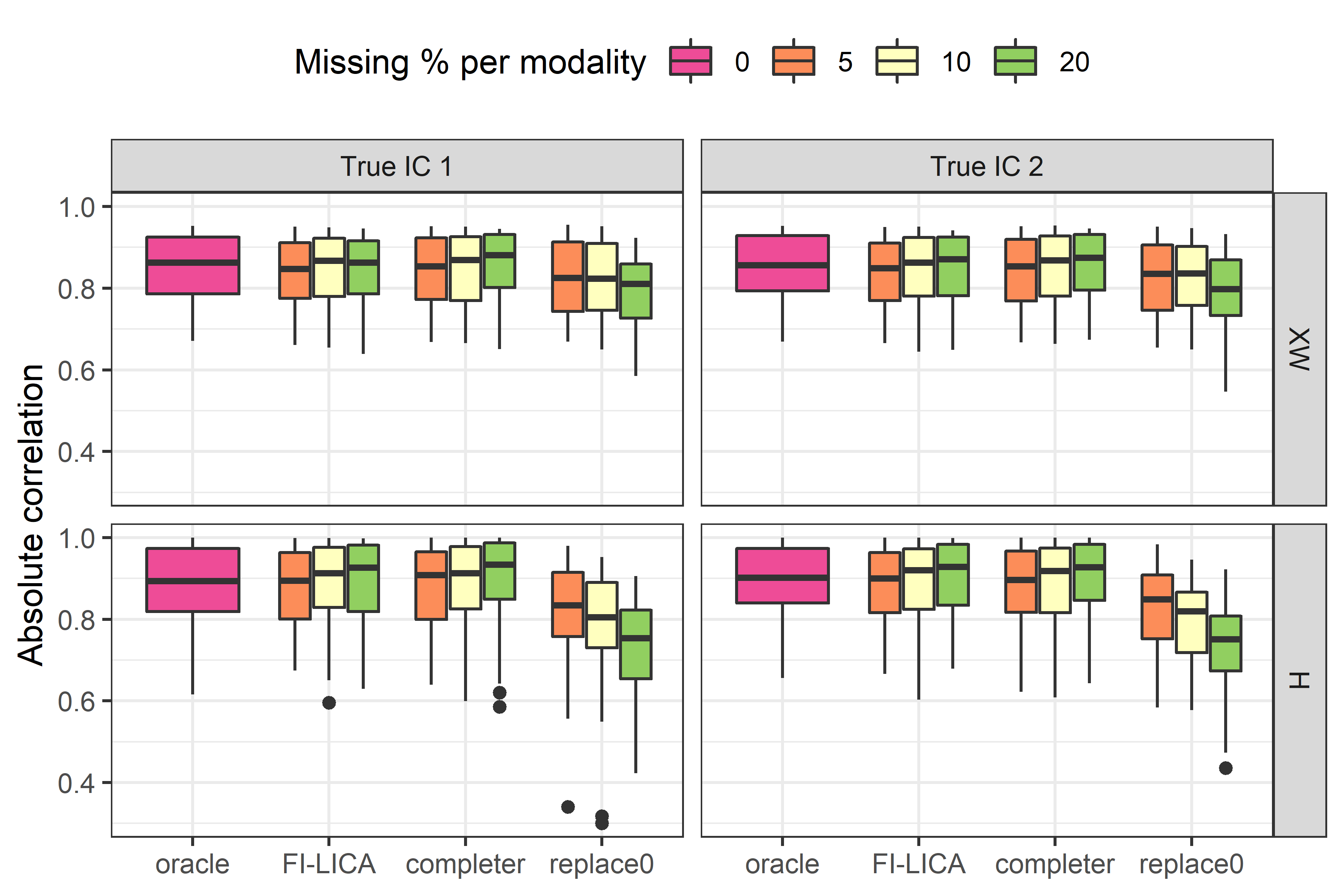}
\end{center}
\caption{Simulation results under the MCAR setting. Box-plots of the absolute correlations between the true $\mathbf{X_W}$ and the best-matching $\mathbf{X_W}$ and box-plots of the absolute correlations between the true $\mathbf{H}$ and the best-matching $\mathbf{H}$, separated by true components and methods. \label{fig:MCAR}}
\end{figure}

Figure \ref{fig:MCAR} shows the box-plots of the absolute correlations between the true $\mathbf{X_W}$ and the best-matching $\mathbf{X_W}$, and the absolute correlations between the true $\mathbf{H}$ and the best-matching $\mathbf{H}$, separated by the two true components, for each of the four comparing methods (FI-LICA, completer, replace0, and the oracle method). 

It can be observed that, under the MCAR setting, although replace0 seems to work fine to identify the spatial independent components ($\mathbf{X_W}$) when the missing percentage is small, FI-LICA starts to outperform replace0 when the missing percentage becomes large. In addition, FI-LICA outperforms replace0 in recovering the shared subject loadings ($\mathbf{H}$), especially with a large percentage of missing. 
The performance of FI-LICA appears comparable to the complete case analysis. However, one major limitation of the complete case analysis is that the shared subject loadings $\mathbf{H}$ are only estimated for subjects present in all modalities -- with 100 full subjects and two modalities, the complete case analysis will end up losing 40\% of data if there is 20\% of missing per modality -- and there is no way we can recover the missing $\mathbf{H}$ values for subjects with missing data using the complete case analysis.

\subsubsection{Simulation setting 2: MAR (Continuous H)}

\begin{figure}
\begin{center}
A. \includegraphics{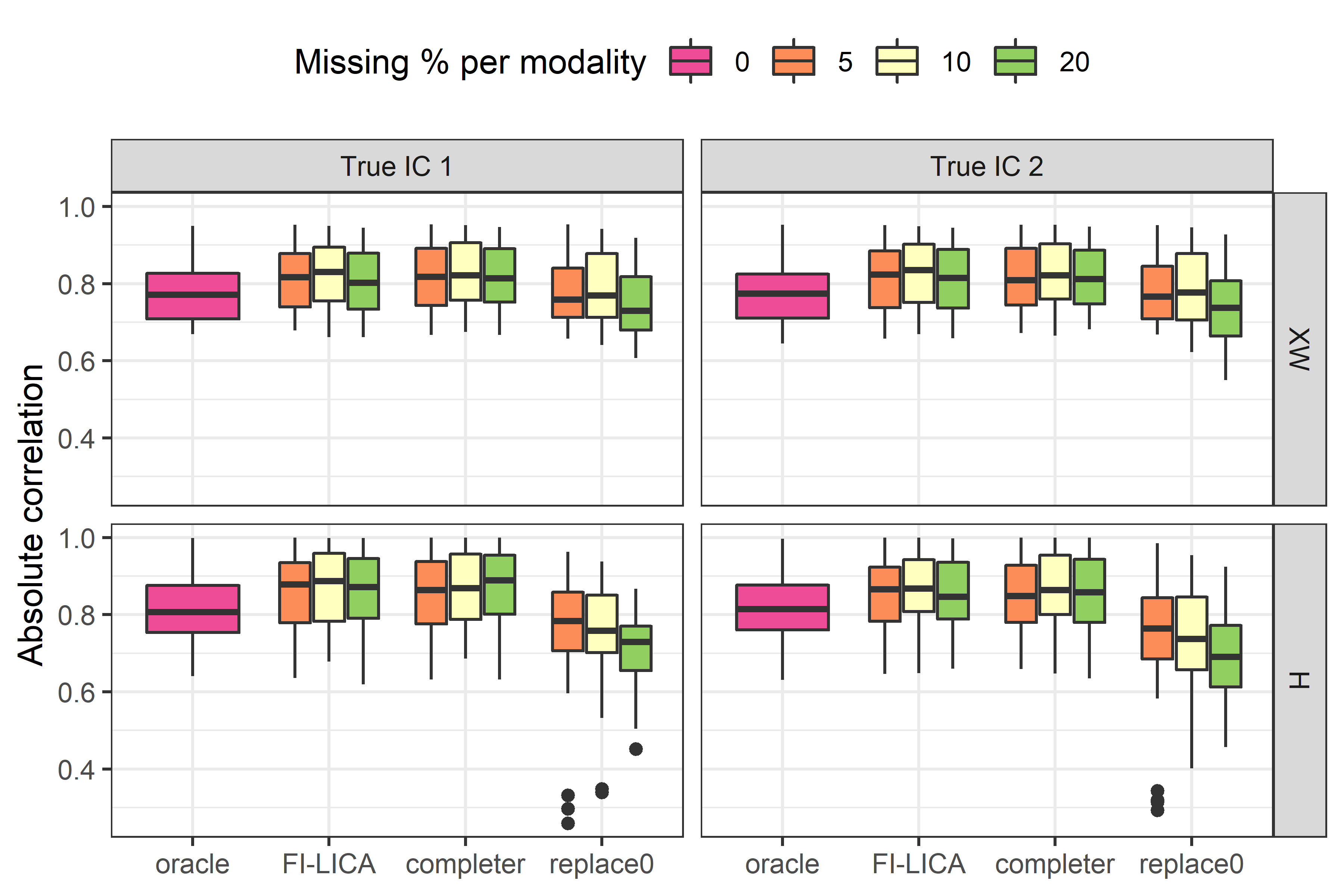} \\
B. \includegraphics[scale=0.93]{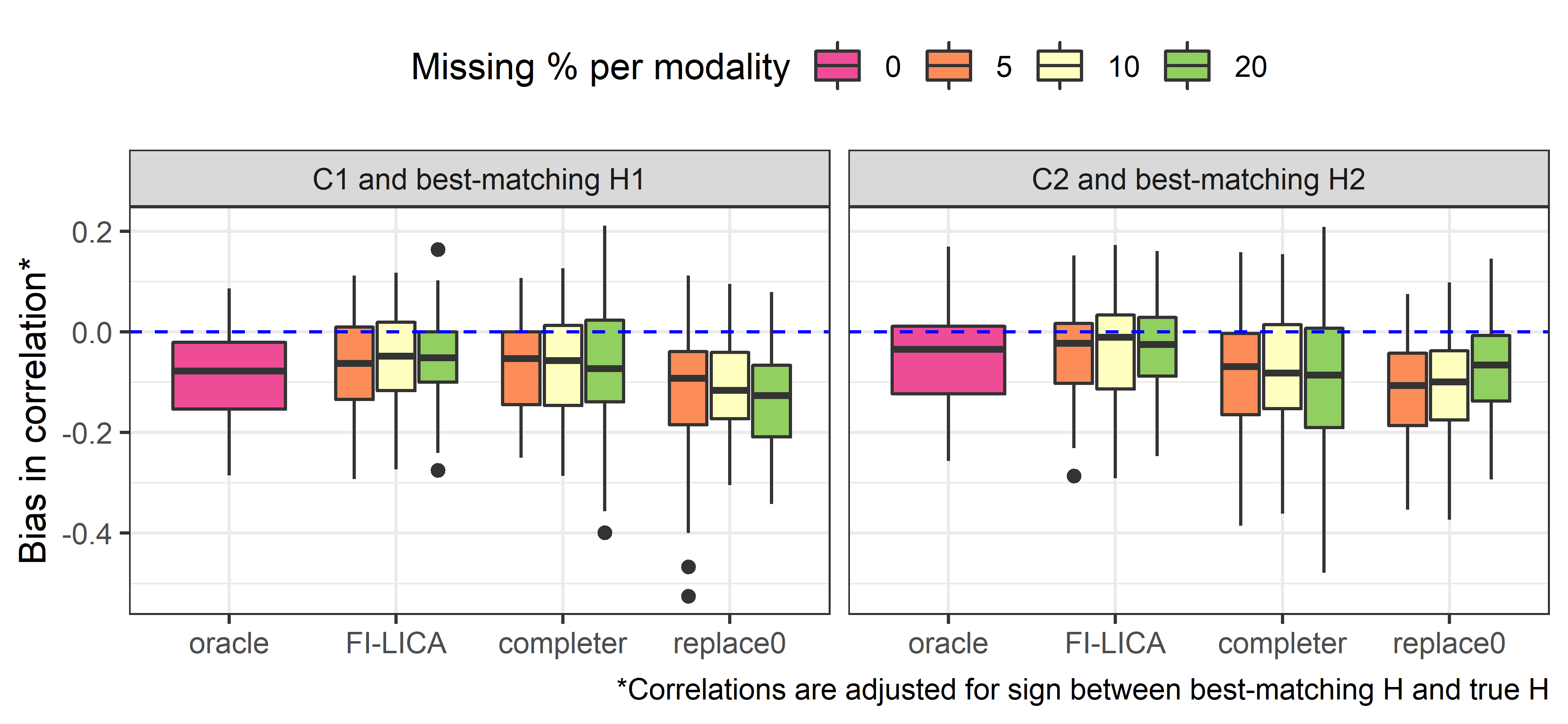}
\end{center}
\caption{Simulation results under the MAR setting with continuous $\mathbf{H}$. 
A: Box-plots of the absolute correlations between the best-matching $\mathbf{X_W}$ and the true $\mathbf{X_W}$, and box-plots of the absolute correlations between the best-matching $\mathbf{H}$ and the true $\mathbf{H}$, separated by true components and methods. 
B: Box-plots of the biases of the correlations between $C_1$ and the best-matching $H_1$, and box-plots of the biases of the correlations between $C_2$ and the best-matching ${H_2}$, separated by methods. 
\label{fig:MAR_ctnsH}}
\end{figure}

Figure \ref{fig:MAR_ctnsH}A shows the same type of box-plots as in Figure \ref{fig:MCAR} but under the MAR setting with continuous $\mathbf{H}$. 
It shows that FI-LICA outperforms replace0 in estimating both the spatial maps ($\mathbf{X_W}$) and the shared subject loadings ($\mathbf{H}$), especially for $\mathbf{H}$ when there is a large percentage of missing. 
Complete case analysis performs similarly to FI-LICA, but the latent $\mathbf{H}$ information for subjects with missing data is not available and cannot be recovered in complete case analysis. 

Figure \ref{fig:MAR_ctnsH}B shows the box-plots of the biases of the correlations between $C_1$ and best-matching $H_1$ with respect to the truth (0.5), and the box-plots of the biases of the correlations between $C_2$ and best-matching $H_2$ with respect to the truth (0.3), separated by methods.
It can be observed that the biases of FI-LICA are the closest to zero, compared to the two current practices (completer and replace0), with complete case analysis exhibiting large variations especially when the missing percentage is large. This indicates that FI-LICA is not only able to recover the missing information of $\mathbf{H}$ for subjects with missing but also able to preserve in its recovered $\mathbf{H}$ the underlying relationship with the variables associated with the missing mechanism.

\subsubsection{Simulation setting 3: MAR (Mixed H)}

\begin{figure}
\begin{center}
A. \includegraphics{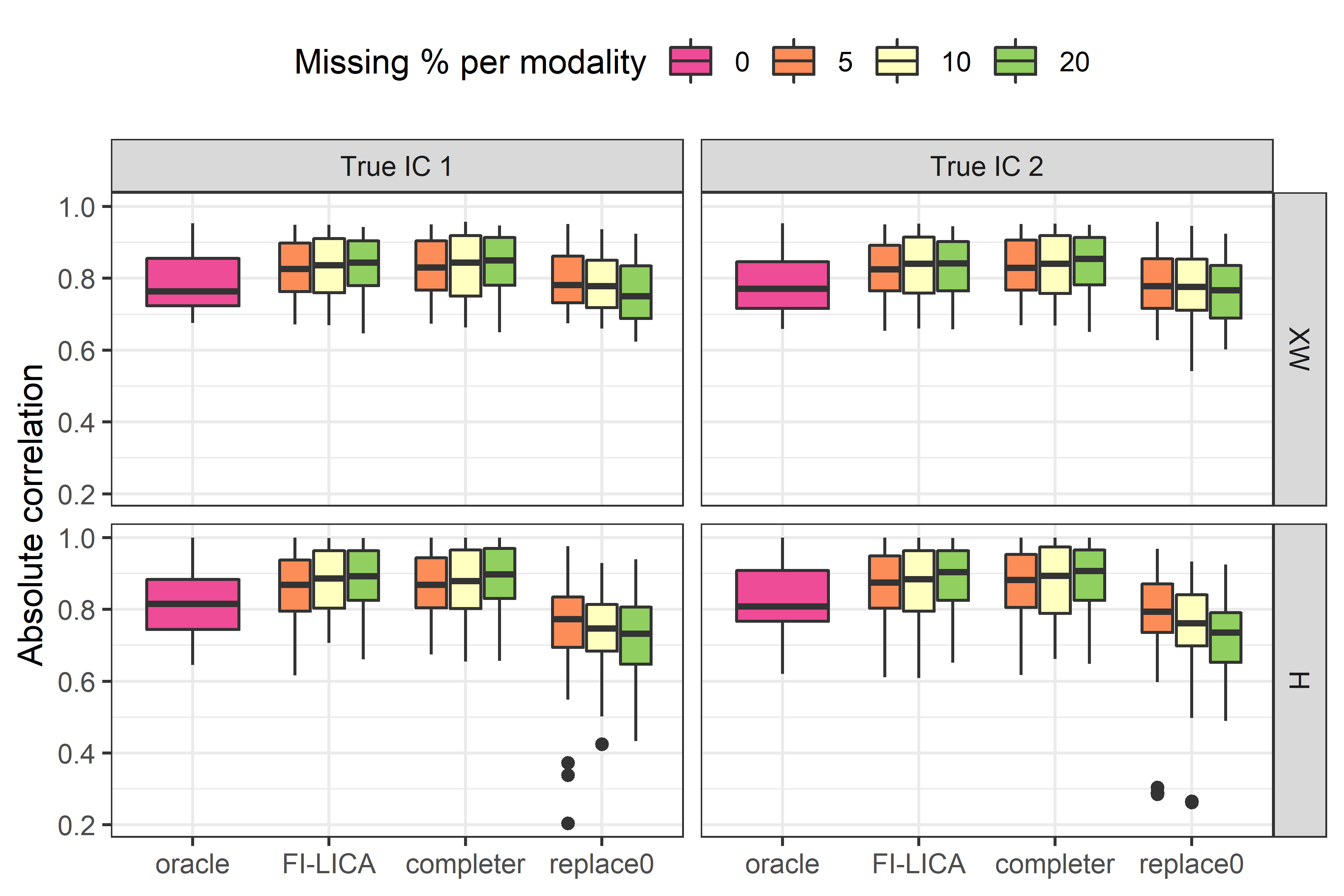} \\
B. \includegraphics[scale=0.93]{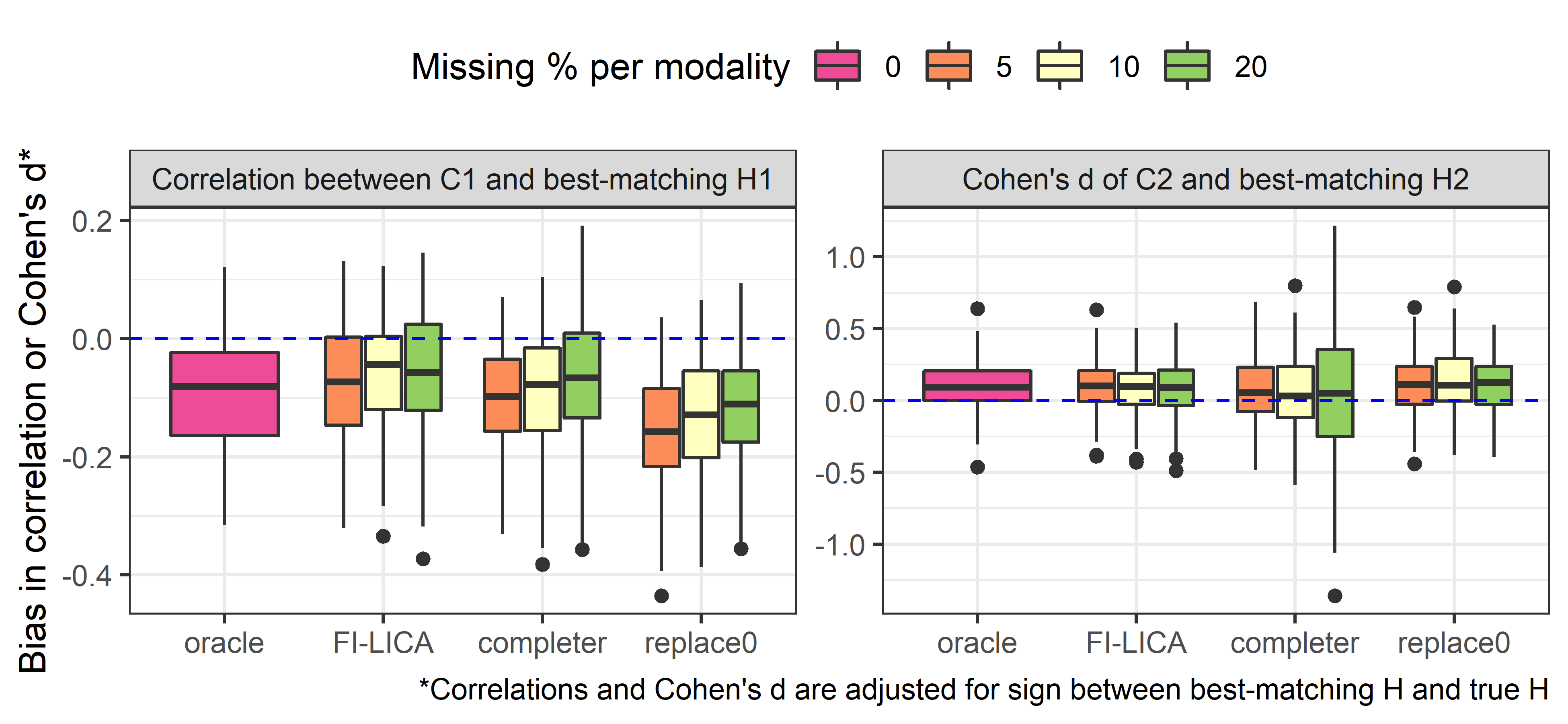}
\end{center}
\caption{Simulation results under the MAR setting with mixed $\mathbf{H}$. 
A: Box-plots of the absolute correlations between the best-matching $\mathbf{X_W}$ and the true $\mathbf{X_W}$, and box-plots of the absolute correlations between the best-matching $\mathbf{H}$ and the true $\mathbf{H}$, separated by true components and methods. 
B: Box-plots of the biases of the correlations between $C_1$ and the best-matching $H_1$, and box-plots of the biases of the Cohen's d values between $C_2$ and the best-matching ${H_2}$, separated by methods. 
\label{fig:MAR_mixedH}}
\end{figure}

Figure \ref{fig:MAR_mixedH}A shows the same type of box-plots as in MAR setting with continuous $\mathbf{H}$ (Figure \ref{fig:MAR_ctnsH}A) but under the MAR setting with mixed $\mathbf{H}$. Similar patterns are observed as in the MAR setting with continuous $\mathbf{H}$.
FI-LICA outperforms replace0 in estimating both the independent components ($\mathbf{X_W}$) and the shared subject loadings ($\mathbf{H}$), especially for $\mathbf{H}$ with a large percentage of missing. 
Complete case analysis performs similarly to FI-LICA, but the latent $\mathbf{H}$ information for subjects with missing data is not available and cannot be recovered in complete case analysis.

Figure \ref{fig:MAR_mixedH}B shows the box-plots of the biases of the correlations between $C_1$ and the best-matching $H_1$ with respect to the truth (0.5), and the biases of the Cohen's d values between $C_2$ and the best-matching ${H_2}$ with respect to the truth (0.5), for each of the methods. 
It can be observed that the biases of FI-LICA are the closest to zero, compared to the two current practices (completer and replace0). This indicates that even when the distributional assumption of $\mathbf{H}$ is violated, FI-LICA is not only able to recover the missing information of $\mathbf{H}$ for subjects with missing but also is robust to preserving in its recovered $\mathbf{H}$ the underlying relationship with the variables associated with the missing mechanism.

\section{Data application}
\label{sec:analysis}

\subsection{Participants}

We applied our proposed algorithm FI-LICA to the human brain multimodal imaging data from the Alzheimer’s Disease Neuroimaging Initiative (ADNI), a longitudinal multicenter study designed for the early detection and tracking of Alzheimer’s disease. 
Specifically, we aim to investigate 1) whether the estimated subject score $\mathbf{H}$ can be used to classify the current diagnostics of the participants -- cognitively normal (CN) vs Alzheimer’s Disease (AD), and 2) whether it can be used to predict AD transition for participants with mild cognitive impairment (MCI). 

The data were downloaded from the ADNI database (\url{http://adni.loni.usc.edu}). The initial phase (ADNI-1) recruited 800 participants, including approximately 200 healthy controls, 400 patients with late MCI, and 200 patients clinically diagnosed with AD over 50 sites across the United States and Canada and followed up at 6- to 12-month intervals for 2-3 years. ADNI has been followed by ADNI-GO and ADNI-2 for existing participants and enrolled additional individuals, including early MCI. To be classified as MCI in ADNI, a subject needed an inclusive Mini-Mental State Examination score of between 24 and 30, subjective memory complaint, objective evidence of impaired memory calculated by scores of the Wechsler Memory Scale Logical Memory II adjusted for education, a score of 0.5 on the Global Clinical Dementia Rating, absence of significant confounding conditions such as current major depression, normal or near-normal daily activities, and absence of clinical dementia.
All studies were approved by their respective institutional review boards and all subjects or their surrogates provided informed consent compliant with HIPAA regulations.

%\subsection{Sample size}

This analysis used three modalities -- T1, AV45 PET, and FDG PET. Each PET modality was matched with one T1 data within one year of its scan date. Data with the earliest available PET scan was considered as baseline. We focused on 1291 subjects who had no missing in T1 but may have missing data in AV45 PET or FDG PET at baseline. We noticed that the two PET modalities might be matched with different T1 scans. To ensure that each subject had only one baseline measure, we took the earlier T1 data if the two T1 dates were within 6 months of each other, and if the two T1 dates were more than 6 months apart, we took the earlier T1 data along with its matched PET data and excluded the other pair of matched T1 and PET data. The Venn diagram of the sample size within the three modalities is displayed in Figure \ref{fig:adni_mod_venn_diagram}. 
Participants' characteristics are provided in Table \ref{tab:tabtwo}.

\begin{table}[h]
\centering \footnotesize
\caption{Participants' characteristics at baseline (ADNI Dataset)} \label{tab:tabtwo} 
%\begin{tabular}{@{}p{2.2cm}@{} @{}p{2cm}@{} @{}p{2cm}@{} @{}p{2cm}@{} @{}p{2.2cm}@{} @{}p{2cm}@{} @{}p{2cm}@{}}
\vspace{3mm}
\begin{tabular}{lrrrrrr}
\hline
  \textbf{Characteristic} & \textbf{CN} & \textbf{MCI} & \textbf{AD} & \textbf{Overall} \\
  & (Cognitively Normal) & (Mild Cognitive Impaired) & (Alzheimer’s Disease) &  \\
  & N = 451 & N = 627 & N = 213 & N = 1,291 \\
\hline
\textbf{Age at scan}: mean (SD) & 73.2 (6.9) & 72.2 (7.8) & 75.0 (8.3) & 73.0 (7.6) \\
\textbf{Sex}: n (\%) & & & &  \\
~~~Female & 252 (55.9\%) & 277 (44.2\%) & 92 (43.2\%) & 621 (48.1\%) \\ 
~~~Male & 199 (44.1\%) & 350 (55.8\%) & 121 (56.8\%) & 670 (51.9\%) \\
\textbf{Ethnicity}: n (\%) & & & & &  \\
~~~Hispanic & 23 (5.1\%) & 25 (4.0\%) & 11 (5.2\%) & 59 (4.6\%) \\ 
~~~Non-Hispanic & 425 (94.2\%) & 600 (95.7\%) & 201 (94.4\%) & 1,226 (95.0\%) \\
~~~Unknown & 3 (0.7\%) & 2 (0.3\%) & 1 (0.5\%) & 6 (0.5\%) \\
\textbf{Race}: n (\%) & & & & &  \\
~~~White & 395 (87.6\%) & 580 (92.5\%) & 196 (92.0\%) & 1,171 (90.7\%) \\
~~~Non-White & 55 (12.2\%) & 43 (6.9\%) & 17 (8.0\%) & 115 (8.9\%) \\ 
~~~Unknown & 1 (0.2\%) & 4 (0.6\%) & 0 (0.0\%) & 5 (0.4\%) \\
\textbf{AD transition status}: n (\%) & & & & &  \\
~~~Dement & 13 (2.9\%) & 160 (25.5\%) & - & 173 (16.0\%) \\
~~~No & 425 (94.2\%) & 438 (69.9\%) & - & 863 (80.1\%) \\ 
~~~Unknown & 13 (2.9\%) & 29 (4.6\%) & - & 42 (3.9\%) \\
\hline
\end{tabular}
\end{table}

\subsection{Imaging processing}

T1-weighted MR images, AV45 PET images, and FDG PET images were downloaded from ADNI repository (https://adni.loni.usc.edu). Cross-sectional image processing was performed using FreeSurfer Version 7. Vertex-level cortical thickness was obtained from T1-weighted MR images on fsaverage5 space with 6mm FWHM smoothing. The preprocessed PET images were downloaded from the ADNI website. Either six five-minute frames (ADNI1) or four five-minute frames (ADNI GO/2) are acquired 30 to 60 minutes post-injection. Separate frames were extracted from the image file and co-registered to the first extracted frame of the raw image file (frame acquired at 30-35 min post-injection). The base frame image and the co-registered frames were recombined into a co-registered dynamic image set. The images were further processed using PET surfer \citep{greve2014cortical} with partial volume correction with 6mm extent, and the vertex-level data were sampled on fsaverage5 space with 6mm FWHM smoothing. The vertex dimension in all three modalities is 20,484 (10,242 vertices each in the left and right hemispheres).

\subsection{Approaches} 

FI-LICA, replace0, and completer analysis were applied to the three modalities. 40 components were used for estimation in all three methods. 10,000 LICA iterations with 20 updates were used for our proposed method FI-LICA. For the replace0 analysis, since LICA was unstable and broke as early as the 17th iteration, 16 LICA iterations were used. For complete case analysis, 10,000 LICA iterations were used. 

In the classification analysis of CN and AD participants, for each of the three methods, we fitted the penalized logistic regression with $\mathbf{H}$ (40 subject scores) as well as gender and age at the scan. 
10-fold cross-validation was used to determine the optimal model. Note that both FI-LICA and replace0 used full sample size whereas complete case analysis used those available in all three modalities. For fair comparison among the three methods, we also ran FI-LICA and replace0 using the complete cases only. Results of the AUC with its standard error (SE) are presented in Table \ref{tab:data_penalize_all}. 

In the analysis of predicting AD transition for MCI participants, for each of the three methods, we fitted the penalized Cox proportional hazards model with $\mathbf{H}$ (40 subject scores) as well as gender and age at the scan. 
10-fold cross-validation was used to determine the optimal model. Again, for fair comparison among the three methods, we ran FI-LICA and replace0 using the complete cases only. Results of the C-index with its standard error (SE) are presented in Table \ref{tab:data_penalize_all}.

\subsection{Results}

In terms of the classification performance of participants' current diagnostics (AD vs CN), our results show that with the full sample size, FI-LICA can classify AD patients better than replace0. When focusing on the completers only, FI-LICA still performs slightly better than replace0, and they both outperform complete case analysis. 
For the AD transition prediction of MCI participants, it can be observed that with the full sample size, FI-LICA performs better than replace0. 
When focusing on the completers only, FI-LICA still has a better performance than replace0 and has a comparably good performance as the complete case analysis.

\begin{table}[] \footnotesize
\caption{Results of AUC and C-index from the data analysis (penalize all)}
\vspace{3mm}
\label{tab:data_penalize_all}
\begin{tabular}{llrrlrr}
\hline
 &  & \multicolumn{2}{c}{\textbf{\begin{tabular}[c]{@{}c@{}}(1) AD/CN classification\\ AUC (SE)\end{tabular}}} &  & \multicolumn{2}{c}{\textbf{\begin{tabular}[c]{@{}c@{}}(2) AD transition prediction\\ C-index (SE)\end{tabular}}} \\ \hline
 &  & \multicolumn{1}{c}{\textit{\begin{tabular}[c]{@{}c@{}}Full sample \\ (n=664)\end{tabular}}} & \multicolumn{1}{c}{\textit{\begin{tabular}[c]{@{}c@{}}Completers \\ (n=473)\end{tabular}}} &  & \multicolumn{1}{c}{\textit{\begin{tabular}[c]{@{}c@{}}Full sample \\ (n=598)\end{tabular}}} & \multicolumn{1}{c}{\textit{\begin{tabular}[c]{@{}c@{}}Completers \\ (n=498)\end{tabular}}} \\
\textbf{FI-LICA} &  & 0.972 (0.0067) & 0.985 (0.0062) &  & 0.827 (0.018) & 0.830 (0.025) \\
\textbf{Replacing missing with 0} &  & 0.970 (0.0072) & 0.984 (0.0064) &  & 0.798 (0.023) & 0.824 (0.028) \\
\textbf{Complete case analysis} &  & -- & 0.978 (0.0078) &  & -- & 0.830 (0.026) \\ \hline
\end{tabular}
\end{table}

We also performed a sensitivity analysis by penalizing $\mathbf{H}$ but not the covariates gender and age for both analyses and consistent results were obtained. AUC and C-index, as well as their SE, are provided in the supplementary Table \ref{tab:data_not_penalize_covariates}.

\section{Conclusion and Discussion}
\label{sec:conc}

In this paper, we proposed the Full Information Linked ICA algorithm (FI-LICA) to address the missing data problem during multimodal fusion, under the LICA framework. 
We evaluated the performance of our algorithm, as well as two current practices, under two mechanisms of missing data -- MCAR and MAR. 
Our simulation results revealed that replacing missing with zero, one current practice of dealing with missing data in LICA, works fine if we are only interested in estimating the spatial maps $\mathbf{X_W}$ and if the missing percentage is small, but it is not likely to perform well to estimate the latent space matrix $\mathbf{H}$ and use it for future analysis. The complete case analysis, another current practice, works fine to estimate $\mathbf{X_W}$ and $\mathbf{H}$, but there is no way we can obtain the missing information of $\mathbf{H}$ for the excluded subjects. By contrast, our proposed method FI-LICA performed well in estimating both $\mathbf{X_W}$ and $\mathbf{H}$. Further, it is not only able to recover the missing information of $\mathbf{H}$ but also able to preserve the underlying relationship between its recovered $\mathbf{H}$ and covariates that are related to the missing mechanism. 
To illustrate our algorithm, we also applied our method to real-world human brain imaging data to investigate whether the estimated subject score $\mathbf{H}$ can be used for the classification of the current diagnostics (AD vs CN) of participants and the prediction of AD transition of MCI participants. Our method demonstrated better performance in both analyses, compared to the two current practices.

%\section*{Acknowledgments}

\section*{Disclosure statement}

The authors report there are no competing interests to declare.

\section*{Funding}
This work was supported by NIH R01AG062578 (PI: Lee).
Data collection and sharing for this project was funded by the Alzheimer’s Disease Neuroimaging Initiative (ADNI) (National Institutes of Health Grant U01 AG024904) and DOD ADNI (Department of Defense award number W81XWH-12-2-0012). ADNI is funded by the National Institute on Aging, the National Institute of Biomedical Imaging and Bioengineering, and through generous contributions from the following: AbbVie, Alzheimer’s Association; Alzheimer’s Drug Discovery Foundation; Araclon Biotech; Bio-Clinica, Inc.; Biogen; Bristol-Myers Squibb Company; CereSpir, Inc.; Cogstate; Eisai Inc.; Elan Pharmaceuticals, Inc.; Eli Lilly and Company; EuroImmun; F. Hoffmann-La Roche Ltd and its affiliated company Genentech, Inc.; Fujirebio; GE Healthcare; IXICO Ltd.; Janssen Alzheimer Immunotherapy Research \& Development, LLC.; Johnson \& Johnson Pharmaceutical Research \& Development LLC.; Lumosity; Lundbeck; Merck \& Co., Inc.; Meso Scale Diagnostics, LLC.; NeuroRx Research; Neurotrack Technologies; Novartis Pharmaceuticals Corporation; Pfizer Inc.; Piramal Imaging; Servier; Takeda Pharmaceutical Company; and Transition Therapeutics. The Canadian Institutes of Health Research is providing funds to support ADNI clinical sites in Canada. Private sector contributions are facilitated by the Foundation for the National Institutes of Health (www.fnih.org). The grantee organization is the Northern California Institute for Research and Education, and the study is coordinated by the Alzheimer’s Therapeutic Research Institute at the University of Southern California. ADNI data are disseminated by the Laboratory for Neuro Imaging at the University of Southern California.

\bigskip
\begin{center}
{\large\bf SUPPLEMENTARY MATERIAL}
\end{center}

\begin{description}

\item[Sensitivity analysis:] 

Table \ref{tab:data_not_penalize_covariates} provides AUC and C-index, as well as their SE, from the sensitivity analysis where we penalize $\mathbf{H}$ but not the covariates gender and age.

\begin{table}[h] \footnotesize
\caption{Results of AUC and C-index from the data analysis (not penalize covariates)}
\label{tab:data_not_penalize_covariates}
\vspace{3mm}
\begin{tabular}{llrrlrr}
\hline
 &  & \multicolumn{2}{c}{\textbf{\begin{tabular}[c]{@{}c@{}}(1) AD/CN classification\\ AUC (SE)\end{tabular}}} &  & \multicolumn{2}{c}{\textbf{\begin{tabular}[c]{@{}c@{}}(2) AD transition prediction\\ C-index (SE)\end{tabular}}} \\ \hline
 &  & \multicolumn{1}{c}{\textit{\begin{tabular}[c]{@{}c@{}}Full sample \\ (n=664)\end{tabular}}} & \multicolumn{1}{c}{\textit{\begin{tabular}[c]{@{}c@{}}Completers \\ (n=473)\end{tabular}}} &  & \multicolumn{1}{c}{\textit{\begin{tabular}[c]{@{}c@{}}Full sample \\ (n=598)\end{tabular}}} & \multicolumn{1}{c}{\textit{\begin{tabular}[c]{@{}c@{}}Completers \\ (n=498)\end{tabular}}} \\
\textbf{FI-LICA} &  & 0.974 (0.0061) & 0.985 (0.0061) &  & 0.824 (0.018) & 0.828 (0.025) \\
\textbf{Replacing missing with 0} &  & 0.971 (0.0072) & 0.985 (0.0062) &  & 0.795 (0.021) & 0.822 (0.028) \\
\textbf{Complete case analysis} &  & -- & 0.977 (0.0076) &  & -- & 0.826 (0.026) \\ \hline
\end{tabular}
\end{table}

%Note: overall N=1078 for ad transition 

\item[R-package for FI-LICA:] 
The FILICA R package and the simulation experiment data example are available at \url{https://github.com/ruiyangli1/FILICA}. 

\item[ADNI data set:] Data used in the preparation of this article were obtained from the Alzheimer's Disease Neuroimaging Initiative (ADNI) database (\url{adni.loni.usc.edu}), under the data use agreement by ADNI.

\end{description}

\bibliographystyle{unsrtnat}
\bibliography{references}   

\end{document}